\documentstyle[aps,preprint]{revtex}
\tightenlines

\begin{document}
\author{G. G. Cabrera\cite{address}}
\address{Instituto de F\'{\i }sica `Gleb Wataghin',\\
Universidade Estadual de Campinas (UNICAMP),\\
C. P. 6165, Campinas 13.083-970 SP, Brazil\\
and }
\author{N. Garc\'{\i }a}
\address{Laboratorio de F\'{\i}sica de Sistemas Peque\~{n}os y Nanotecnolog\'{\i}a,\\
Consejo Superior de Investigaciones Cient\'{\i}ficas (CSIC),\\
Serrano 144, E-28006 Madrid, Spain}
\title{Low Voltage I-V Characteristics in Magnetic Tunnel Junctions}
\date{November, 2000}
\maketitle

\begin{abstract}
We show that elastic currents that take into account variations of the
tunnel transmitivity with voltage and a large ratio of majority to minority
spin densities of states of the $s$ band, can account for the low voltage
current anomalies observed in magnet-oxide-magnet junctions. The anomalies
can be positive, negative or have a mixed form, depending of the position of
the Fermi level in the $s$ band, in agreement with observations. Magnon
contribution is negligible small to account for the sharp drop of the
magnetoresistance with the voltage bias.
\end{abstract}

\newpage Tunneling of electrons in metal-insulator-metal junctions is an old
phenomenon studied from a long time ago\cite{old,simmons}. However, it is
only quite recently that spin-dependent tunneling between two ferromagnetic
metals has been shown to produce the magnetoresistance effect observed in
those systems\cite{mr,zhang}. In $3d$ ferromagnets, most of the spin
polarization comes from the $d$ bands, while tunneling currents are
dominated by $s$ band contributions. This is so, because $d$ wave functions
are more localized and their effective tunneling barrier is higher\cite
{nico1}. For $Ni$, it has been estimated that the tunneling probability of
the $s$ electrons is of the order of $100-1000$ times that of the $d$
electrons, thus leading to a positive spin polarization in $Ni$ field
emission experiments\cite{meservey}. In the context of tunneling
experiments, the large magnetoresistance effect (25-30 \%) found in \cite
{mr,zhang} is puzzling, since it points to a large polarization of the $s$
band, with a ratio of the densities of states for majority $\left(
N^{(M)}(E)\right) $ and minority $\left( N^{(m)}(E)\right) $ electrons at
the Fermi level $\left( E_{F}\right) $ of the order of 
\begin{equation}
N^{(M)}(E_{F})/N^{(m)}(E_{F})\approx 2.0-2.5\quad ,  \label{ratio}
\end{equation}
in apparent contradiction with energy band calculations for ferromagnetic
metals\cite{moruzzi}.

In addition, a remarkable dependence of the junction conductance with the
voltage bias $\left( V\right) $ has been observed at low voltages (of the
order of a few hundred millivolts). As usual in magnetoresistance
experiments, one compares the resistances for the cases where the
magnetizations at the electrodes are anti-parallel (AP) and parallel (P). In
several experiments reported in Ref.\cite{mr,zhang}, the junction resistance
drops significantly with the applied voltage, with a peak at zero bias
(called {\em zero-bias anomaly) }that is more pronounced for the AP
alignment. The effect is also temperature dependent, the peak being less
sharp at room temperature. Finally, it is found that the junction
magnetoresistance (JMR) has a large decrease with voltages, up to 60\% at
0.5 V in some cases\cite{mr}. It has been argued that this effect can be
attributed to the excitation of internal degrees of freedom by hot electrons
(even at liquid He temperature). Scattering from surface magnons has been
proposed as a mechanism to randomize the tunneling process and open the
spin-flip channels that leads eventually to a sharp drop of the MR\cite
{zhang}. However, this explanation is controversial, since magnon scattering
cross sections are negligibly small to account for such a big drop of
resistance and no spin-flip events have been observed in experiments with
polarized injected electrons in tunnelling phenomena\cite{spinflip}. Also,
the theory given in Ref. \cite{zhang} \underline{assumes tunneling
transmitivities independent of the applied voltages}, and uses a
perturbation scheme only valid for voltages smaller than $\sim 40$ $mV$,
while the data extend to $\sim 400\ mV$.

In the present Letter, we show that the variations of the conductance with
the voltage bias can be simply accounted for by the lowering of the barrier
height with voltages, as given by the Simmons' tunneling theory\cite{simmons}%
. The structure at zero bias is obtained, when one properly takes into
account variations of the density of states with the bias at both magnetic
electrodes. Assuming that the tunneling current comes from the $s$ band, we
formulate a simple model with a parabolic dispersion (free-electron{\em \ }%
like). We obtain different behaviors for the zero-bias anomaly, whether the
Fermi level is located near the bottom ({\em peak}) or top of the band ({\em %
dip}). Fitting with the experiments\cite{mr,zhang} can only be obtained if
one assumes a large spin polarization corresponding to relation (\ref{ratio}%
).

In order to develope our calculation, one has to rewrite Simmons' formulae
with the conductance current written in the form 
\begin{equation}
J^{\left( C\right) }(V)=A\sum_{\sigma ,\mu }%
\displaystyle \int %
\limits_{-\infty }^{\infty }\ dE\ T\left( E,\Delta s,\phi ,V\right) \
N_{L}^{(\sigma )}(E)N_{R}^{(\mu )}(E+V)\left[ f_{L}\left( E\right)
-f_{R}(E+V)\right] \ ,  \label{current}
\end{equation}
where $T\left( E,\Delta s,\phi ,V\right) $ is the transmitivity through the
barrier for energy $E$, parametrized with the mean barrier height $\phi $
and width $\Delta s$\cite{simmons}, the index $C=P,AP$ refers to the
magnetic configuration (parallel or anti-parallel), and $\ N_{L,R}$ and $%
f_{L,R}$ are the densities of states and the Fermi distributions for the
left and right electrodes, respectively. In ferromagnets, one has to
distinguish between {\em majority }$(M)$ and {\em minority }$(m)${\em \ }%
spin bands and the super-indices in the densities of states and in the sum
in expression (\ref{current}) label the allowed processes for spin
conserving tunneling, for both magnetic configuration, $P$ and $AP$. For
parallel alignment, the factor of the densities of states that enter in (\ref
{current}) is 
\begin{equation}
N_{L}^{(m)}(E)N_{R}^{(m)}(E+V)+N_{L}^{(M)}(E)N_{R}^{(M)}(E+V)\ ,  \label{pc}
\end{equation}
while for the anti-parallel configuration, where majority and minority are
interchanged for the left and right electrodes, one has to consider 
\begin{equation}
N_{L}^{(m)}(E)N_{R}^{(M)}(E+V)+N_{L}^{(M)}(E)N_{R}^{(m)}(E+V)\ .  \label{apc}
\end{equation}
Concerning equation (\ref{current}), several remarks are in order.

\begin{enumerate}
\item[{\em i)}]  In his original treatment of the tunneling problem\cite
{simmons}, Simmons considers the case of very flat conduction bands for the
metal electrodes and takes the densities of states as constants. However,
for $s$ bands the density of states varies as the square root of the energy,
and for magnetic junctions this cannot be neglected, especially near the
band edges, where the variation is bigger. Zero-bias anomalies in normal
non-magnetic metals has been previously reported, in cases where the
structure of the density of state is important\cite{wyatt}.

\item[{\em ii)}]  Expression (\ref{current}) involves an integral over all
energies, but states that are deep in the band are cut off exponentially by
the tunneling probability. As a net result, the conductance is dominated by
electrons that are near the Fermi level, and (\ref{current}) approximately
factorizes in the form 
\begin{equation}
J^{\left( C\right) }(V)\approx \left( \sum_{\sigma ,\mu =m,M}^{C}N^{(\sigma
)}(E_{F})N^{(\mu )}(E_{F}+V)\right) J^{\left( S\right) }(V)\
=D^{(C)}(E_{F},V)\ J^{\left( S\right) }(V)  \label{factor}
\end{equation}
where $J^{\left( S\right) }(V)$ is the Simmons' tunneling current as a
function of the voltage bias and 
\begin{equation}
D^{(C)}(E_{F},V)=\sum_{\sigma ,\mu =m,M}^{C}N^{(\sigma )}(E_{F})N^{(\mu
)}(E_{F}+V)\ .  \label{density}
\end{equation}
In (\ref{factor}), we are assuming that both electrodes are made from the
same ferromagnetic metal. The term $J^{\left( S\right) }(V)$ is the Simmons'
contribution, is spin independent and carries all the information concerning
the tunneling barrier. As shown in \cite{simmons}, it has no quadratic term
in the voltage for small bias, and no zero-bias anomaly.
\end{enumerate}

In Fig. 1, we show the variation with voltage of the Simmons resistance for
typical barriers, with the resistance normalized at zero bias. A large
variation is observed in all the examples, but the resistance has no peak or
dip at zero voltage. Except for the structure at zero bias, the overall
variation of the Simmons' resistance is of the order of what is observed in
experiments (or even may vary faster with voltage in some cases). Some
experimental results are also shown for comparison.

Next, we introduce the factor $D^{(C)}(E_{F},V)$, defined in (\ref{density}%
), in the conductance calculation. We model the density of states of the $s$
bands with a parabolic dependence (free-electron like) in the form 
\[
N^{(\sigma )}(E)=\frac{\Omega }{4\pi ^{2}}\left( \frac{2m_{e}}{\hbar ^{2}}%
\right) ^{3/2}\sqrt{\pm \left( E-E_{\sigma }\right) },\qquad \sigma =m,M, 
\]
where $\Omega $ is the volume of the sample (electrode), $m_{e}$ is the
electron mass, and the $\pm $ sign refers to the cases where we are in the
bottom or in the top of the conduction band, respectively. In formulating
the Stoner model within a naive band theory, $\left| E_{m}-E_{M}\right| $
should yield the exchange of the $s$ band. But Fermi surfaces of transition
metals are very intricate, with contributions from electron and hole-like
carriers and with different shapes for majority and minority spin sheets. In
this context, $E_{m}$ and $E_{M}$ come from the band structure and $\Delta
E=\left| E_{m}-E_{M}\right| $ may be very different from the true exchange
of the band.

To parametrized our results, and denoting by $E_{F}$ the Fermi energy, we
define 
\[
\begin{array}{l}
E_{F}^{M}\equiv \left| E_{F}-E_{M}\right| , \\ 
E_{F}^{m}\equiv \left| E_{F}-E_{m}\right| , \\ 
E_{F}^{M}\equiv \lambda \ E_{F}^{m},\quad \lambda >1,
\end{array}
\]
which includes both cases, bottom and top of the band. The ratio of the
densities of states at the Fermi level is given by $%
N_{L}^{(M)}(E_{F})/N_{L}^{(m)}(E_{F})=\sqrt{\lambda }$. Several
possibilities can be realized, wether majority and minority carriers are
electrons or holes. When both are electrons or holes, the factors $%
D^{(C)}(E_{F},V)$ can be expanded in series in $V$, yielding a linear term
in $V$ that is responsible for the zero-bias anomaly: 
\[
\begin{array}{l}
D_{\pm }^{(P)}(V)\approx \left( \left[ N^{(m)}(E_{F})\right] ^{2}+\left[
N^{(M)}(E_{F})\right] ^{2}\right) \left( 1\pm p^{(P)}\left| V\right| \right)
, \\ 
\\ 
D_{\pm }^{(AP)}(V)\approx \left( 2N^{(m)}(E_{F})N^{(M)}(E_{F})\right) \left(
1\pm p^{(AP)}\left| V\right| \right) ,
\end{array}
\]
where the $\pm $ sign labels the bottom and top cases respectively, with the
slopes of the linear terms given by 
\[
\begin{array}{l}
p^{(P)}=%
{\displaystyle {1 \over E_{F}^{m}\left( 1+\lambda \right) }}%
\ , \\ 
\\ 
p^{(AP)}=%
{\displaystyle {\lambda +1 \over 4\lambda E_{F}^{m}}}%
\ .
\end{array}
\]
When we have a mixed case, {\em i.e. }one of the spin is electron-like and
the other hole-like, no linear term appears in $D^{(P)}(V)$. On the other
hand, for $D^{(AP)}(V)$, the slope of the linear term is given by 
\[
p^{(AP)}=\mp \left( 
{\displaystyle {\lambda -1 \over 4\lambda E_{F}^{m}}}%
\right) \ ,
\]
where the $-$ ($+$) sign applies when the majority carriers are electrons
(holes). In Fig. 2, we display results of our calculation for examples of
typical barriers. The value of the magnetoresistance at zero bias was taken
from Ref. \cite{zhang}, with 
\[
N_{L}^{(M)}(E_{F})/N_{L}^{(m)}(E_{F})=\sqrt{\lambda }\approx 2.2\quad .
\]
In Fig. 2 {\em a)}, we show the case when the Fermi level is in the bottom
of the $s$ band, with a linear decrease of the resistance with the voltage
bias for both magnetic configurations ($AP$ and $P$). If the Fermi level is
in the top of both spin bands, we initially get a linear increase of the
resistance which, after some voltage value, is dominated by the Simmons'
term. This case is displayed in part {\em c)} of Fig. 2. In Fig. 2 {\em b)},
we display the situation where the majority band ($\uparrow $) is almost
filled (holes) and the minority ($\downarrow $) is almost empty (electrons).
The resistance for the $P$ setup, exhibits no linear term. In Fig. 2 {\em a)}%
, we also show experimental results taken from Ref. \cite{zhang}. We have
not tried an optimum fitting with experiments, but it is clear that
experimental results can only be explained assuming a large polarization of
the $s$ band. Note that the insets in Fig. 2 {\em a)}-2 {\em c)} sketch the
band configurations for both spins.

The change in tunnel resistance or magnetoresistance (MR) is given by 
\begin{equation}
\frac{\Delta R}{R}=\frac{R_{AP}-R_{P}}{R_{AP}}\quad ,  \label{mr}
\end{equation}
where again, $AP$ and $P$ refer to the magnetic configuration of the
ferromagnetic electrodes. This ratio, as it is evident from relation (\ref
{factor}), is almost independent of the Simmons' term, not depending on
details of the tunneling process. In Fig. 3{\em \ A)}, we display results of 
$\Delta R/R$ corresponding to the examples of Fig. 2. In {\em B)}, we take
different experimental results found in the literature\cite{mr}. Note that
when the Fermi level lies near the top of the band, there is an increase of
the MR. Eventually, we may reach the minority spin band edge, with a
vanishing density of states, for which 
\[
R_{AP}\rightarrow \infty \quad . 
\]

Temperature $(T)$ effects can also be taken into account through relation (%
\ref{current}), with the broadening of the Fermi distributions, but a rough
estimation shows that the effect should be similar to that of an applied
voltage $V\approx 2T$, with an effective lowering of the barrier height, a
smaller resistance, and the softening of the zero-bias anomaly, in agreement
with experiments.

From our calculations presented above the following conclusions are
pertinent:

\begin{enumerate}
\item[i)]  The overall variation of the tunnel current with voltage \cite
{mr,zhang} can be explained by elastic tunneling using the well known
Simmons' formula\cite{simmons} and is due to the lowering of the barrier by
the applied voltage. This is at variance with the calculations in Ref. \cite
{zhang}, where they argue that this effect is negligible. Therefore, magnons
are not needed to explain the experiments;

\item[ii)]  The anomalies in the currents and the magnetoresistances can be
explained within this simple framework, provided that the ratio of majority
spin to minority spin electrons is of the order of $2.2-2.5$, for the data
of Ref.\cite{mr,zhang}. If one is allowed to choose the adequate
configuration of the s bands (see Fig.2), a maximum, a minimum or a mix of
both can appear at the anomaly (as it has been observed in Ref.\cite{sharma}%
);

\item[iii)]  From band structures calculations \cite{moruzzi}, it is not
clear to us that the above polarization of the $s$ band can be justified.
There may be other oxidation states inside the metal, at the interface, and
in the oxide layer, that contribute to the polarization of the current;

\item[iv)]  Alternatively, it may also happen, as it has been suggested in
Ref.\cite{nico1,berko,ivan}, that the current is dominated by conduction
paths that provide large values of magnetoresistance\cite{nico2} due to
domain wall scattering\cite{wall}, and then there is also contribution of $d$%
-electrons. In this case, the density of states will have mixed
contributions from $s$ and $d$-electrons, with a variety of topologies in
the MR\cite{nico3};

\item[v)]  The main conclusion is that the magnetoresistance is a mapping of
the spin up and down densities of states in the metals and the barrier and
cannot be assigned only to the bulk ferromagnetic metals, and many mixing
possibilities exist for explaining the physical measurements.\bigskip 
\end{enumerate}

{\bf Acknowledgments.} GGC acknowledges partial support from Brazilian
FAPESP {\em (Funda\c{c}\~{a}o de Amparo \`{a} Pesquisa do Estado de S\~{a}o
Paulo) }and CNPq {\em (Conselho Nacional de Desenvolvimento Cient\'{\i }fico
e Tecnol\'{o}gico)}.

\newpage

\begin{center}
{\bf FIGURE CAPTIONS}
\end{center}

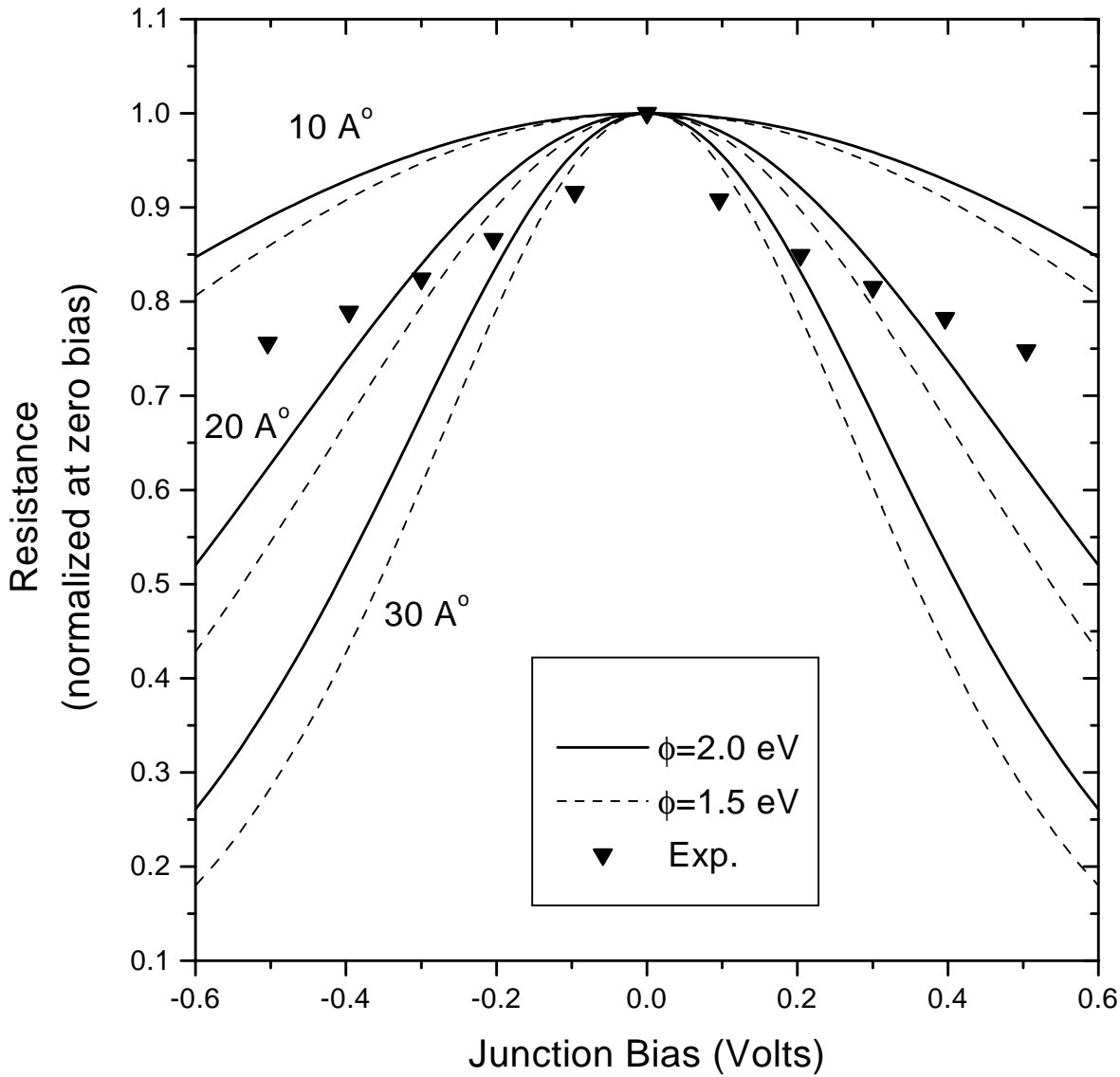
{\bf Fig. 1 }Variation of the Simmons' resistance with voltage for several
tunnel barriers. Data is normalized at zero bias. Experimental results from 
\cite{zhang} are also shown (solid triangles) as a reference.

\bigskip

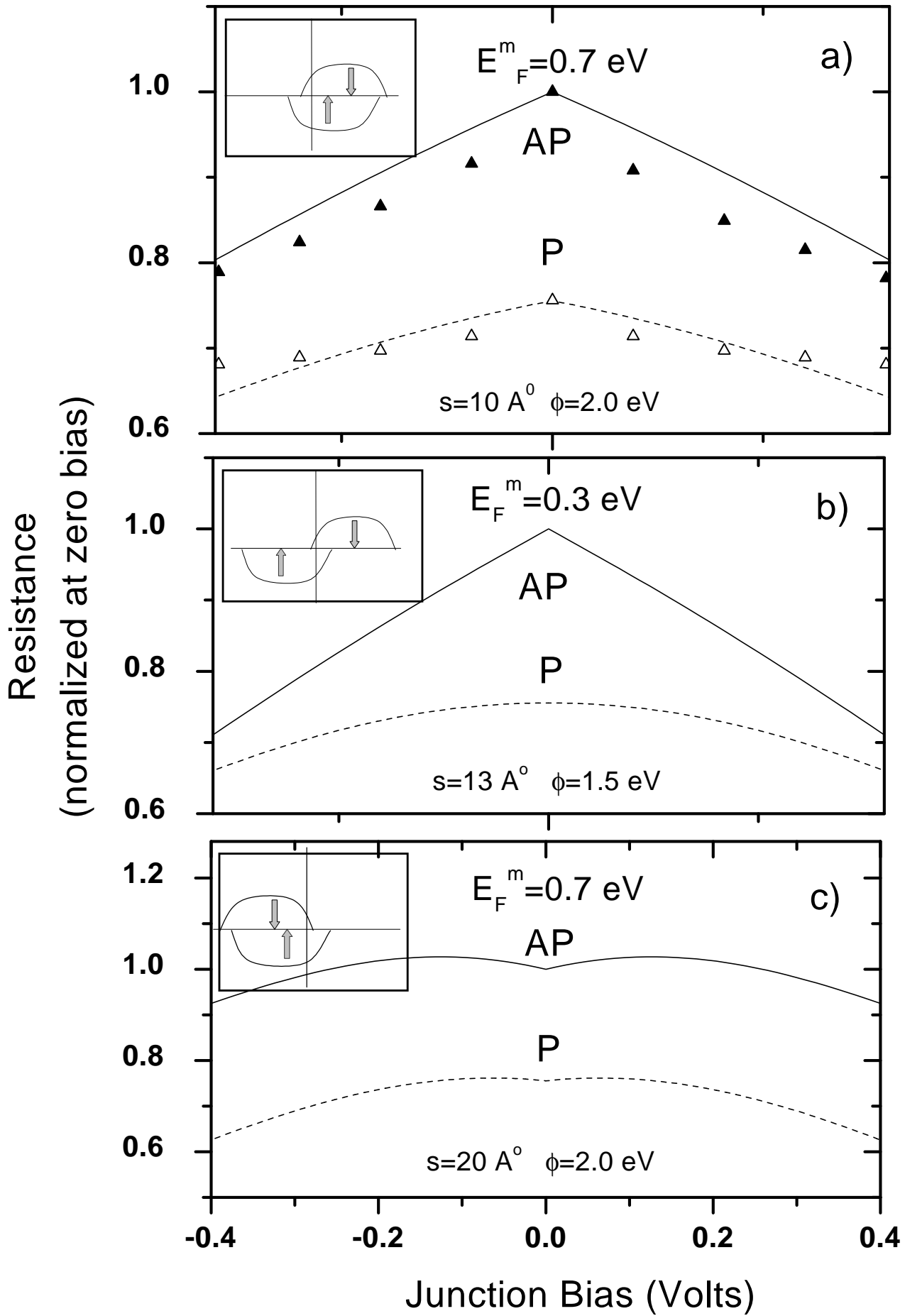
{\bf Fig. 2 }Resistance as a function of the voltage bias for the two
configurations of the magnetic electrodes and for different $s$ band
structures (they are shown in the insets). Parameters for the tunneling
barriers are given in each figure. Spin $\uparrow $ is taken as the majority
band in all cases. As a reference, experimental results take from \cite
{zhang} are shown in part {\em a)}, where a good agreement with our
calculation is obtained.

\bigskip

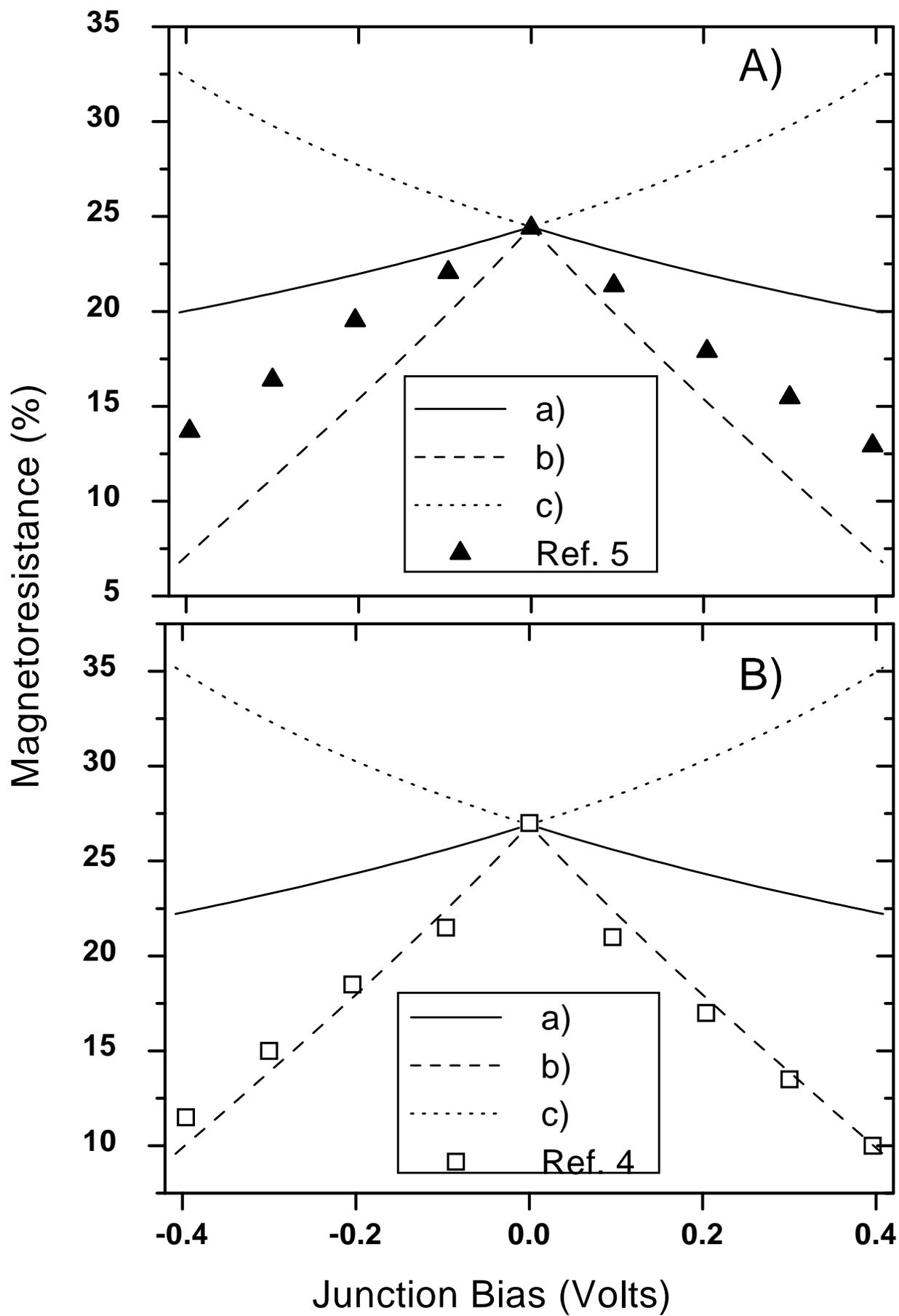
{\bf Fig. 3 }Magnetoresistance, as defined in (\ref{mr}), for all the cases
depicted in Fig. 2. Densities of states are adjusted at the zero bias value.
In {\em A)},{\em \ }we compare with results from \cite{zhang}, while part 
{\em B)} compares with Ref.\cite{mr}.


\begin{references}
\bibitem{zhang}  S. Zhang, P. M. Levy, A. C. Marley, and S. S. P. Parkin,
Phys. Rev. Lett. {\bf 79}, 3744 (1997). This paper uses perturbation theory
in $\left( eV/\hbar \omega _{c}\right) $, where $\hbar \omega _{c}\approx
100\ meV$, is roughly the maximum magnon energy, and the calculation should
be valid when this ratio is much smaller than one. However, the theory is
extrapolated to values of $V\sim 400\ meV$, where it is clearly not valid.
Also, for this range, the assumption made that the values of the tunneling
probability are voltage independent is not feasible, as can be easily
deduced from Simmons' tunneling theory\cite{simmons} and as shown in Fig. 1
of this paper. No justification whatsoever is given for the extremely high
magnon scattering cross section. This is just another adjustable parameter
of the theory.
\end{references}
\end{document}